\documentclass[aps,prl,amsmath,amssymb,superscriptaddress, twocolumn]{revtex4}

\usepackage{graphicx}

\begin{document}

\title{Broadband Coherent Enhancement of Transmission and Absorption in Disordered Media}

\author{Chia Wei Hsu}
\affiliation{Department of Applied Physics, Yale University, New Haven, Connecticut 06520, USA}
\email{chiawei.hsu@yale.edu}

\author{Arthur Goetschy}
\affiliation{Laboratoire Mat\'{e}riaux et Ph\'{e}nom\`{e}nes Quantiques, Universit\'{e} Paris Diderot and CNRS F-75205 Paris, France}

\author{Yaron Bromberg}
\affiliation{Department of Applied Physics, Yale University, New Haven, Connecticut 06520, USA}

\author{A. Douglas Stone}
\affiliation{Department of Applied Physics, Yale University, New Haven, Connecticut 06520, USA}

\author{Hui Cao}
\affiliation{Department of Applied Physics, Yale University, New Haven, Connecticut 06520, USA}

\begin{abstract}
We study the optimal diffusive transmission and absorption of broadband or polychromatic light in a disordered medium.
By introducing matrices describing broadband transmission and reflection, we formulate an extremal eigenvalue problem where the optimal input wavefront is given by the corresponding eigenvector.
We show analytically that a single wavefront can exhibit strongly enhanced total transmission or total absorption across a bandwidth that is orders of magnitude broader than the spectral correlation width of the medium, due to long-range correlations in coherent diffusion.
We find excellent agreement between the analytic theory and numerical simulations.
\end{abstract}

\maketitle

One exciting development in optics in recent years is the coherent control of diffusing light in a disordered medium by shaping input wavefronts using a spatial light modulator (SLM)~\cite{2012_Mosk_nphoton, 2015_Horstmeyer_nphoton}. 
Initially the emphasis was on using wavefront shaping (WFS) to focus light onto a wavelength-scale region (speckle) behind or within the disordered medium~\cite{2007_Vellekoop_OL, 2008_Vellekoop_PRL}, with potential applications for imaging; the optimal input wavefront in this case can be found by a simple sequential optimization of each pixel on the SLM, since each contributes to the local field at the focal spot independently.
More recently there has been progress in the more challenging problem of optimizing {\it global} properties of the fields, such as the total transmitted power through the medium~\cite{2012_Kim_nphoton, 2014_Popoff_PRL, 2014_Gerardin_PRL}.
Motivation came from theoretical concepts first formulated in the context of mesoscopic electron transport and localization theory~\cite{1984_Dorokhov_SSC, 1986_Imry_EPL, 1988_Mello_AP, 1994_Nazarov_PRL}, where it was predicted that in a lossless diffusive medium there would always exist sample-specific ``open channels"
that will be transmitted almost perfectly.
A closely related effect is the coherent enhancement of absorption (CEA) to near unity via WFS in a disordered medium that on average only absorbs a small fraction of the input light~\cite{2011_Chong_PRL,2015_Liew_CEA}.
Incomplete control of the input wavefronts reduces the possible enhancements~\cite{2013_Goetschy_PRL, 2013_Yu_PRL}, but large enhancements are still observable under realistic conditions~\cite{2014_Popoff_PRL,2015_Liew_CEA}.

The physical basis of these coherent control effects is manipulation of the multiple-scattering interference in the medium to violate the expected behavior for incoherent diffusion.
Hence these effects would seem to be intrinsically narrowband, limiting their applications in contexts such as power delivery, communications, or energy conversion, in which larger bandwidths may be required. 
The expected bandwidth is limited by the frequency correlation scale, $\delta \omega $, which for lossless transmission is the inverse of the time to diffuse across the thickness $L$ of the medium,  $\delta \omega \approx D/L^2$ ($D = l c/d$ is the diffusion constant in $d$ dimensions, and $l$ is the transport mean free path)~\cite{Akkermans_book}; for CEA, $\delta \omega \approx c/l_a$, where $l_a$ is the ballistic absorption length~\cite{2011_Chong_PRL}.
For a broadband signal with bandwidth $\Delta \omega \gg \delta \omega$, a natural hypothesis is that the effective number of independent frequencies would be $M_{\rm eff} \approx 1+ \Delta \omega / \delta \omega$, and that the maximal achievable enhancement decreases as $1/M_{\rm eff}$.
Indeed this is exactly the behavior found in experiments maximizing the focal intensity of polychromatic light on a single speckle spot using SLMs~\cite{2011_vanBeijnum_OL, 2012_Small_OL, 2013_Paudel_OE, 2015_Andreoli_srep}.
However we will show that this is not the case for the total transmission or absorption due to the long-range spectral correlations of coherent diffusion~\cite{1987_Stephen_PRL, 1988_Feng_PRL, 1989_Pnini_PRB, 1995_Rogozkin_PRB}, 
 which are unimportant for speckle statistics but play a major role for the global properties~\cite{2014_Popoff_PRL}.
These correlations dramatically reduce the effective number of independent degrees of freedom, and instead of the linear scaling, we find $M_{\rm eff} \approx \sqrt{\Delta \omega / \delta \omega}$, allowing substantial coherent control of transmission and absorption over large bandwidths.
For example, for a lossless diffusive sample with average 2\% transmission, the total transmission can be enhanced 10 times across bandwidth $\Delta \omega \approx 60 \delta \omega$;
similarly for a thick diffusive sample with average 3\% absorption, the total absorption can be enhanced 10 times across $\Delta \omega \approx 60 \delta \omega$.

\begin{figure}[t]
   \centering
   \includegraphics{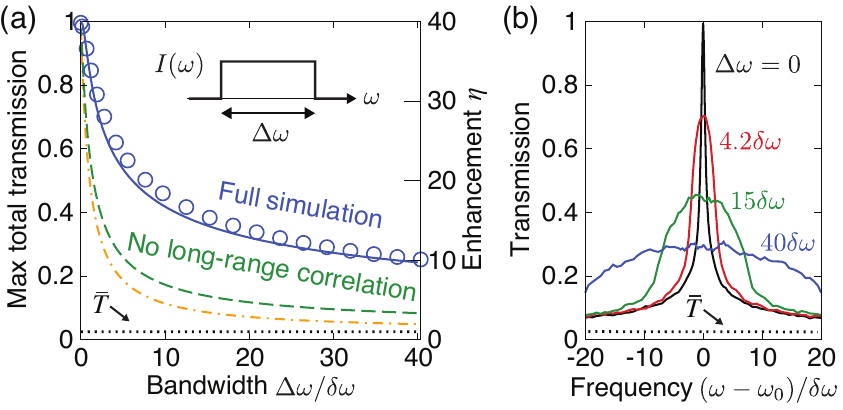} 
   \caption{Broadband transmission open channels.
   (a) Maximal eigenvalue $T_{\rm max}$ of the broadband flux matrix $A$ and the enhancement $\eta = T_{\rm max}/\bar{T}$ obtained from numerical simulations (blue circles) and analytic theory accounting for long-range correlation (blue line), showing the highest achievable frequency-integrated transmission across bandwidth $\Delta \omega$.
   The two lines below show the would-be maximal transmission (green, dashed) and the transmission of the monochromatic open channel (orange, dot-dashed) if there were $1+\Delta\omega/\delta\omega$ uncorrelated frequencies. 
   Black dotted line indicates the average transmission.
   (b)
   Transmission as a function of frequency when the input wavefront is fixed to the optimal eigenvector with different bandwidths $\Delta \omega$.
   }
   \label{fig:broadband_Tmax}
\end{figure}

We begin by defining a broadband flux matrix, based on the monochromatic transmission matrix $t(\omega)$ that relates the incident field $|\psi_{\rm in}\rangle$ to the transmitted field $|\psi_{\rm t}\rangle = t |\psi_{\rm in}\rangle$; the field vectors are written in the basis of $N$ input and output modes carrying unit flux, and we assume $N \gg 1$.
The monochromatic transmitted flux $ \langle \psi_{\rm t} |\psi_{\rm t}\rangle$ is the expectation value 
$ \langle \psi_{\rm in} | t^\dagger (\omega) t (\omega)|\psi_{\rm in}\rangle$ of the Hermitian matrix $t^\dagger t$; it follows that the most open channel for monochromatic light corresponds to the largest eigenvalue of $t^\dagger t$~\cite{1984_Dorokhov_SSC, 1986_Imry_EPL, 1988_Mello_AP, 1994_Nazarov_PRL, 2011_Choi_PRB, 2012_Shi_PRL, 2014_Liew_PRB, 2014_Pena_ncomms, 2015_Davy_PRL, 2015_Davy_ncomms}.
For polychromatic light, the role of $t^\dagger t$ is replaced by 
\begin{equation}
\label{eq:A_def}
A = \int d\omega I(\omega) t^\dagger(\omega) t(\omega) ,
\end{equation}
where $I(\omega)$ is the power spectrum of the incident light normalized to $\int d\omega I(\omega)  = 1$.
When the transmitted flux is measured with a sufficiently-long integration time, beating between different frequencies averages away,
and the total transmission for incident light with spectrum $I(\omega)$ and wavefront $|\psi_{\rm in}\rangle$ is simply $\langle \psi_{\rm in} | A |\psi_{\rm in}\rangle$.
Since $A$ is still Hermitian, the optimal wavefront is again given by the eigenvector with the largest eigenvalue.
A broadband reflection flux matrix can be defined similarly, with $r^\dagger r$ replacing $t^\dagger t$.
Note that a monochromatic open channel at some frequency within the spectral envelope is generally no longer an eigenvector of $A$, so we immediately know that it will not provide the optimal broadband transmission.
The optimization figure of merit here is essentially the frequency-averaged transmission with the bandwidth and weight specified by $I(\omega)$; as we shall see later in Fig.~\ref{fig:broadband_Tmax}(b), the optimal wavefront has its transmission enhanced rather uniformly across the target bandwidth,
so optimizations aiming for uniformity (such as maximin) will yield similar results.

In the diffusive regime ($\lambda \ll l \ll L$, where $\lambda$ is wavelength),
each matrix $t^\dagger(\omega) t(\omega)$  has a bimodal eigenvalue density $p_{t^\dagger t}(T) = \bar{T}/(2T\sqrt{1-T})$ where $\bar{T}$ is the average transmission~\cite{1984_Dorokhov_SSC, 1986_Imry_EPL, 1988_Mello_AP, 1994_Nazarov_PRL}.
The distribution has support up to $T=1$, meaning monochromatic open channels always exist for $N \gg 1$ in the diffusive regime.
Since the transmission matrices at different frequencies do not commute, the eigenvalue density of the broadband matrix $A$ will be very different, as we show below.

\begin{figure}[tb]
   \centering
   \includegraphics{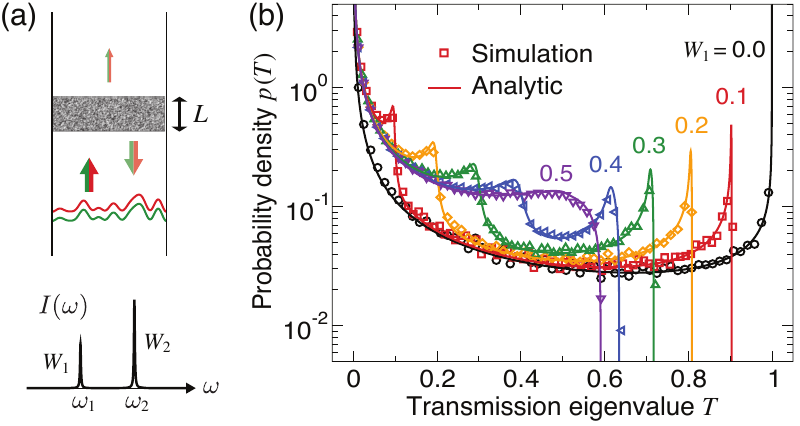} 
   \caption{Total transmission through a disordered slab for incident light with two discrete frequencies.
    (a) Schematic setup, with a disordered slab in a multimode waveguide and polychromatic light incident with a shared wavefront.
    (b) Density of the polychromatic transmission eigenvalues as calculated numerically by solving the wave equation (symbols) and analytically using Eq.~\eqref{eq:gA} from free probability theory (lines). Numbers indicate the intensity weight $W_1$. 
    }
   \label{fig:M2}
\end{figure}

First, we study the simpler situation when $A$ is given by a sum of matrices at discrete frequencies that we assume are so widely separated that correlations between them are negligible. Subsequently we will adapt this theory to treat a continuous input spectrum with long-range correlations, relevant for more experiments.
Hence initially we take $I(\omega) = \sum_{m=1}^M W_m \delta(\omega-\omega_m)$ and assume no correlation between the $M$ matrices, $\{t(\omega_m) \equiv t_m\}$.
The setup for $M=2$ is illustrated in Fig.~\ref{fig:M2}(a).
The eigenvalue density for a sum of large, mutually uncorrelated, non-commuting random matrices can be treated by methods developed in free probability theory, which generalizes the concept of statistical independence to such matrices~\cite{Voiculescu_book, Tulino_book}.
Specifically one can apply an addition rule~\cite{1986_Voiculescu_addition} to find an implicit equation for the eigenvalue density of their sum.
For the matrix $A$, define $g_A(z)$ as the Stieltjes transform (resolvent) of the eigenvalue density $p_A$; applying the addition rule, 
one finds that the unknown resolvent $g_A$ can be obtained from the following implicit equation (details in~\cite{note_SI})
\begin{equation}
\label{eq:gA}
z + \frac{M-1}{g_A(z)} = \sum_{m=1}^M W_m g_{t_m^\dagger t_m}^{-1}(W_m g_A(z)),
\end{equation}
with the known resolvent $g_{t_m^\dagger t_m}$ that is determined from the bimodal distribution $p_{t_m^\dagger t_m}$.
We then apply standard root-finding algorithms to this equation to find $g_A(z)$ and obtain the desired eigenvalue density through the inverse Stieltjes transform $p_A(T) = - \lim_{\epsilon \to 0^+} {\rm Im} g_A(T+i\epsilon)/\pi$.
Results for the general ($M>2$) cases are given in Fig.~S1 in \cite{note_SI}.
Here we examine the simpler $M=2$ case for different combinations of weights $\{W_1, W_2\}$, shown as solid curves in Fig.~\ref{fig:M2}(b) (here, $\bar{T}_1 = 0.027$, $\bar{T}_2 = 0.021$).
The $W_1=0$, $W_2 = 1$ case corresponds to the monochromatic bimodal distribution.
With increasing $W_1$, the upper edge $T_{\rm max}$ decreases as expected;
the residual peaks near $T \approx W_1$ and $T \approx W_2$ can be traced back to the open channels of the constituent matrices $W_1 t_1^\dagger t_1$ and $W_2 t_2^\dagger t_2$.
From the case $W_1=W_2=1/2$, we see that $T_{\rm max} \approx 0.59$ is larger than the $(1+\bar{T})/2 \approx 1/2$ one would obtain from using the monochromatic open channels as the input wavefront.
However, we note that in the limit of $\bar{T}_1, \bar{T}_2 \to 0$, $T_{\rm max}$ will still approach $1/2$.

We perform numerical simulations to validate the analytic prediction.
As illustrated in Fig.~\ref{fig:M2}(a), we simulate a 2D disordered slab of thickness $L$ and width $W=3L$
in a waveguide geometry with background refractive index $n_0 = 1.5$ and slab permittivity $\epsilon({\bf r})$ randomly sampled between $n_0^2 \pm 0.9$ at each grid point.
Using the recursive Green's function method~\cite{1991_Baranger_PRB}, we obtain the $N$-by-$N$ transmission matrix (here $N=647$) of the wave equation $[\nabla^2 + (\omega/c)^2 \epsilon({\bf r})] \psi({\bf r}) = 0$ for 600 realizations of disorder, at two frequencies $\omega_1 = 390 c/L$ and $\omega_2 = 410 c/L$ (average transmissions $\bar{T}_1 = 0.027$, $\bar{T}_2 = 0.021$; the variation of $N$ is negligible) that are much further than $\delta \omega$ apart (here $\omega_2 - \omega_1 \approx 290 \delta \omega$). 
The resulting eigenvalue densities of the two-frequency matrix $A$, shown as symbols in Fig.~\ref{fig:M2}(b), agree perfectly with the analytic prediction with no fitting parameters.

To see the effect of transmission correlations, we perform wave simulations of the diffusive medium for a broadband input with uniform spectral weights $I(\omega)$ over bandwidth $\Delta \omega$, centered at $\omega_0 = 400 c/L$ (where $\bar{T} = 0.025$). 
The numerically obtained maximum eigenvalue $T_{\rm max}$ of the broadband matrix $A$ is plotted as blue circles in Fig.~\ref{fig:broadband_Tmax}(a), as a function of $\Delta \omega / \delta \omega$, where $\delta \omega =  0.069 c/L \approx 21 D/L^2$ is defined as the full width at half maximum (FWHM) of the transmission spectrum for the monochromatic open channel (black line in Fig.~\ref{fig:broadband_Tmax}(b));
note that the FWHM of the open channel coincides with the FWHM of the speckle intensity correlation (see Fig.~S2 in \cite{note_SI}).
In Fig.~\ref{fig:broadband_Tmax}(a), we find that at all bandwidths, $T_{\rm max}$ (blue circles) is much larger than the prediction of the uncorrelated model when the effective number of independent frequencies is taken as $M_{\rm eff} = 1 + \Delta \omega / \delta \omega$ (green dashed line),
which itself is larger than the frequency-averaged transmission of the monochromatic open channel, $(1+\bar{T}\Delta \omega / \delta \omega)/(1+\Delta \omega / \delta \omega)$, when one assumes $M_{\rm eff} = 1 + \Delta \omega / \delta \omega$ (orange dot-dashed line).
The transmission spectra of the optimal broadband eigenvectors cover the target bandwidth rather uniformly, as shown in Fig.~\ref{fig:broadband_Tmax}(b) for representative bandwidths.
Clearly spectral correlations beyond the FWHM are critical and should allow much greater coherent control of broadband transmission.

\begin{figure}[t]
   \centering
   \includegraphics{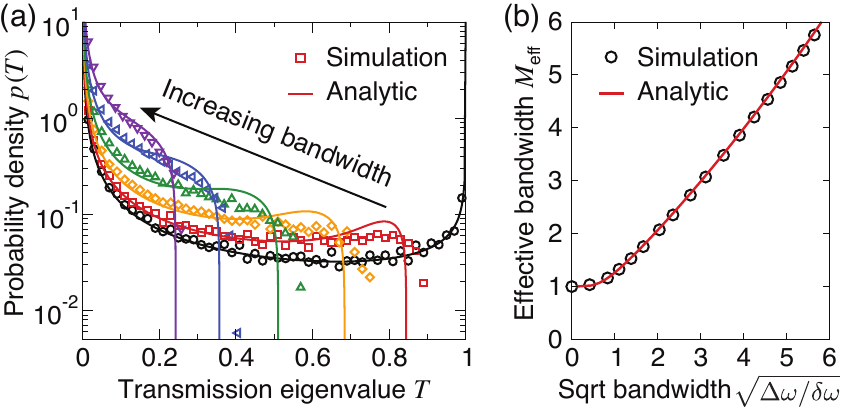} 
   \caption{
   (a) Density of the broadband transmission eigenvalues for various bandwidths, calculated numerically from simulations (symbols; bandwidths $\Delta \omega / \delta \omega =$ 0, 1.0, 2.4, 5.9, 15, 40) and analytically from Eq.~\eqref{eq:gA_equal_weight} with an effective number $M_{\rm eff}$ of independent frequencies (lines; $M_{\rm eff} = 1.0, 1.3, 1.7, 2.4, 3.9, 6.6$).
   (b) $M_{\rm eff}$ as a function of the square-root bandwidth, evaluated numerically from Eq.~\eqref{eq:M_and_var} (symbols) and analytically from Eqs.~\eqref{eq:Meff}-\eqref{eq:C_T} (line).
   }
   \label{fig:broadband_pT_Meff}
\end{figure}

To account for these spectral correlations, we adopt an approach similar to the treatment of spatial correlations in Ref.~\cite{2014_Popoff_PRL}.
We hypothesize that even in the presence of spectral correlations, the eigenvalue density can still be described by Eq.~\eqref{eq:gA}, but with $M$ replaced by some effective number of independent frequencies, $M_{\rm eff} < 1 + \Delta \omega/\delta \omega$.
We focus on the case where the spectral weights $W_m$ is uniform, for which Eq.~\eqref{eq:gA} takes a simpler form
\begin{equation}
\label{eq:gA_equal_weight}
\frac{g_A(z)}{M_{\rm eff}} =  g_{t^\dagger t} \left( z + \frac{M_{\rm eff}-1}{g_A(z)} \right).
\end{equation}
This coincides with Eq.~(3) in Ref.~\cite{2013_Goetschy_PRL} when $A$ is taken to be $\tilde{t}^\dagger \tilde{t}$ with $\tilde{t}$ being a ``filtered" matrix that only has a fraction $m_1 = 1/M_{\rm eff}$ of the input channels (columns) of the full matrix $t$,
as in experiments where all output lights are measured but only a fraction of the incident channels is controlled by the SLM~\cite{2014_Popoff_PRL}.
Given this equivalence, we can use a property of the filtered matrix~\cite{2013_Goetschy_PRL}
\begin{equation}
\label{eq:M_and_var}
\frac{1}{M_{\rm eff}} = \frac{{\rm Var}(\tilde{\tau}) }{{\rm Var}({\tau})},
\end{equation}
to determine $M_{\rm eff}$, where  $\tilde{\tau}$ and $\tau$  are the eigenvalues of $A$ and of $t_m^\dagger t_m$ respectively.
With the broadband eigenvalue density from simulations (symbols in Fig.~\ref{fig:broadband_pT_Meff}(a)), we confirm that Eq.~\eqref{eq:M_and_var} provides the correct value of $M_{\rm eff}$ that, through Eq.~\eqref{eq:gA_equal_weight}, predicts analytical eigenvalue densities (lines in Fig.~\ref{fig:broadband_pT_Meff}(a)) that agree well with the numerical data.
$M_{\rm eff}$ obtained in this manner scales with the square root of the bandwidth (circles in Fig.~\ref{fig:broadband_pT_Meff}(b)).

The quantity ${\rm Var}(\tilde{\tau})$ can be expressed in terms of the disorder average of certain products of four transmission amplitudes $t_{ab}$,
and the disorder averages can be carried out analytically using impurity-averaged perturbation theory (details in~\cite{note_SI} and Figs.~S3-S4). We find
\begin{equation}
\label{eq:Meff}
\frac{1}{M_{\rm eff}} =  \iint \frac{d\omega_1 d\omega_2}{\Delta \omega^2} \frac{C^{(T)}(\omega_1,\omega_2)}{C^{(T)}(\omega_0, \omega_0)},
\end{equation}
where $C^{(T)}(\omega_1,\omega_2) $ is the mean-normalized spectral correlation 
$\langle T_a(\omega_1) T_a(\omega_2) \rangle / \langle T_a(\omega_1)\rangle \langle T_a(\omega_2) \rangle - 1$ of the total transmission $T_a = \sum_b |t_{ba}|^2$, with the brackets denoting average over disordered samples; the dependence on mode index $a$ drops out due to the normalization.
In our system, $C^{(T)}$ is well described by (see Fig.~S4(a) in~\cite{note_SI})
\begin{equation}
\label{eq:C_T}
C^{(T)}(\omega_1, \omega_2) = \frac{1}{N\bar{T}} \left[ \frac{2}{x} \frac{\sinh(x) - \sin(x)}{\cosh(x) - \cos(x)} - \bar{T} \right],
\end{equation}
where $ x = \sqrt{2 {|\omega_1 - \omega_2|}L^2/{D}}$.
In Eq.~\eqref{eq:C_T}, the first term is the long-range correlation that decays as  $|\omega_1-\omega_2|^{-1/2}$ (see Ref.~\cite{1989_Pnini_PRB}),
while the second term is a finite-$\bar{T}$ correction~\cite{2004_Garcia-Martin_PRL, note_SI}.
Eqs.~\eqref{eq:Meff}-\eqref{eq:C_T} provide an analytic expression to calculate $M_{\rm eff}$ without free parameters and perfectly agrees with the $M_{\rm eff}$ obtained from simulations, as shown in Fig.~\ref{fig:broadband_pT_Meff}(b).
The blue solid line in Fig.~\ref{fig:broadband_Tmax}(a) is calculated with this analytic expression of $M_{\rm eff}$,
and it explains the much larger potential transmission enhancement through WFS than expected from the uncorrelated model.
Specifically, when $\Delta \omega$ falls in the regime $1 \ll \sqrt{ \Delta \omega / \delta \omega} \ll 1/\bar{T}$, the relevant values of $C^{(T)}$ are dominated by the $|\omega_1-\omega_2|^{-1/2}$ tail in the long-range contribution, giving rise to the scaling of $M_{\rm eff} \approx \sqrt{ \Delta \omega / \delta \omega}$ and a parametrically larger $T_{\rm max}$. 
Note that Eqs.~\eqref{eq:Meff}-\eqref{eq:C_T} show that $M_{\rm eff}$ and $T_{\rm max}$ depend only on the bandwidth $\Delta \omega$ and the average transmission $\bar{T}$.

In addition to enhancing transmission, coherent superpositions of the input modes can also enhance absorption (CEA).
Consider a thick diffusive scattering medium with $\lambda \ll l \ll \sqrt{l l_a/d} < L$, where $l_a$ is the ballistic absorption length.
As the thickness $L$ is larger than the diffusive absorption length $L_a = \sqrt{l l_a/d}$, the transmitted flux is exponentially small, so any light that is not reflected can be considered absorbed.
As $l \ll l_a$, most incident light is reflected before it propagates far enough to be absorbed, so the average absorption is low.
However, there exist eigenchannels that can be nearly completely absorbed at one frequency when the number of input channels ({\it i.e.} degrees of freedom to be controlled) is large enough that $N^2 l/l_a \gg 1$~\cite{2011_Chong_PRL, note_CEA}.
The minimum reflection (corresponding to the maximum absorption) is the smallest eigenvalue of $r^\dagger r$, and in the $N\to \infty$ limit the monochromatic eigenvalues follow a known bimodal distribution, 
$p_{r^\dagger r}(R) = 2a \sqrt{(1-R)/(aR) - 1} /(\pi(1-R)^2)$,
where $a \equiv l/l_a \ll 1$~\cite{1996_Bruce_JPA, 1996_Beenakker_PRL}.
For broadband light with spectrum $I(\omega)$, we instead look for the eigenvalues of $A$ as defined in Eq.~\eqref{eq:A_def} just with $t(\omega)$ replaced by $r(\omega)$.

\begin{figure}[t]
   \centering
   \includegraphics{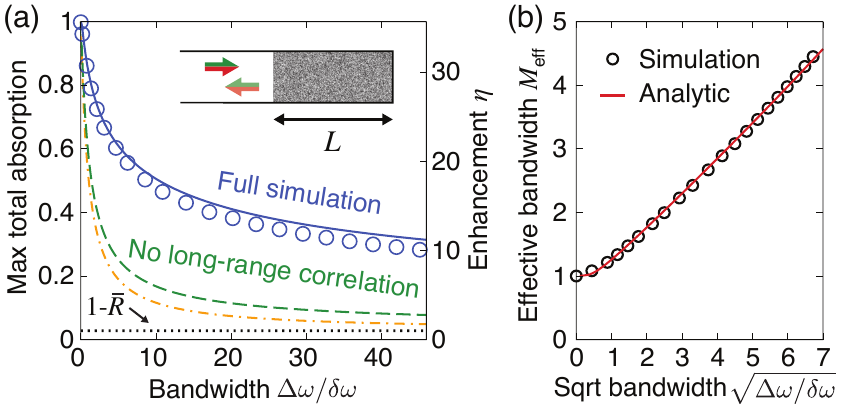} 
   \caption{Broadband coherently enhanced absorption (CEA).
   (a) Maximal frequency-integrated absorption $1-R_{\rm min}$ and the enhancement $\eta = (1-R_{\rm min})/(1-\bar{R})$ obtained numerically (blue circles) and analytically (blue line) across bandwidth $\Delta \omega$.
   The two lines below show the would-be maximal absorption (green, dashed) and the absorption of the monochromatic CEA mode (orange, dot-dashed) if there were $1+\Delta\omega/\delta\omega$ uncorrelated frequencies. 
   Black dotted line indicates the average absorption.
   Inset shows schematic setup of the system; a reflecting boundary on the right ensures that transmission is zero and absorption is $1-R$.
   (b) $M_{\rm eff}$ as a function of the square-root bandwidth, evaluated numerically from Eq.~\eqref{eq:M_and_var} (symbols) and analytically from Eqs.~\eqref{eq:Meff} and \eqref{eq:C_R} (line).
   }
   \label{fig:broadband_CEA}
\end{figure}

We perform numerical simulations for the geometry shown in the inset of Fig.~\ref{fig:broadband_CEA}(a)
with thickness $L$ and width $W=0.43L$
and with a weak uniform absorption ${\rm Im}(\epsilon) = 3 \times 10^{-5}$ in the diffusive medium (corresponding to $a=2\times 10^{-4}$, $\bar{R} = 0.97$, and $N=323$ near $\omega_0 = 1400 c/L$).
Again, we consider broadband incident light with uniform spectral weights $I(\omega)$ over bandwidth $\Delta \omega$, and numerically evaluate the reflection matrices $r(\omega)$ and the eigenvalues of the broadband flux matrix $A$.
The maximum absorption, $1 - R_{\rm min}$, is plotted as blue circles in Fig.~\ref{fig:broadband_CEA}(a) as a function of $\Delta \omega / \delta \omega$,
where $\delta \omega  = 0.14 c/L \approx 12 c/l_a$ is defined as the FWHM of the absorption spectrum for the monochromatic CEA channel (see Fig.~S5(a) in~\cite{note_SI}).
Similar to the broadband lossless transmission, here we find the maximal absorption to be much larger than the prediction if one were to ignore long-range spectral correlation (green dashed and orange dot-dashed lines).

The density of broadband reflection eigenvalues is well described by Eq.~\eqref{eq:gA_equal_weight} with $t^\dagger t$ replaced by $r^\dagger r$ and with $M_{\rm eff}$ given by Eq.~\eqref{eq:M_and_var} (see Fig.~S5(b) in~\cite{note_SI}), confirming the hypothesis that one can use an effective number of independent frequencies to describe the broadband eigenvalue distribution.
Analytically, $M_{\rm eff}$ is again given by Eq.~\eqref{eq:Meff}, just with $C^{(T)}$ replaced by the spectral correlation
$C^{(R)}(\omega_1, \omega_2) \equiv \langle R_a(\omega_1) R_a(\omega_2) \rangle / \langle R_a(\omega_1)\rangle \langle R_a(\omega_2) \rangle - 1$ of the total reflection $R_a = \sum_b |r_{ba}|^2$, which in our system is well described by (see Fig.~S4(b) in~\cite{note_SI})
\begin{equation}
\label{eq:C_R}
C^{(R)}(\omega_1, \omega_2) = \frac{1-\bar{R}}{N\bar{R}(1+\bar{R})} \left[ \frac{2}{1+y} - (1-\bar{R}) \right],
\end{equation}
where $ y={\rm Re}\sqrt{1+ i |\omega_1-\omega_2| l_a /c}$.
Here, the first term decays as $|\omega_1-\omega_2|^{-1/2}$ and is the long-range reflection correlation derived in Ref.~\cite{1995_Rogozkin_PRB},
while the second term is a correction for finite $1-\bar{R}$~\cite{note_SI}.
Eqs.~\eqref{eq:Meff} and \eqref{eq:C_R} provide an analytic expression for $M_{\rm eff}$ and is plotted as the red line in Fig.~\ref{fig:broadband_CEA}(b), with its prediction of the maximal absorption plotted as the blue line in Fig~\ref{fig:broadband_CEA}(a).
When the bandwidth $\Delta \omega$ falls in the regime $1 \ll \sqrt{ \Delta \omega / \delta \omega} \ll 1/(1-\bar{R})$, the $C^{(R)}$ is dominated by the $|\omega_1-\omega_2|^{-1/2}$ tail in the long-range contribution, and $M_{\rm eff}$ scales as $\sqrt{ \Delta \omega / \delta \omega}$, 
giving rise to large potential enhancements of absorption similar to the lossless transmission case.

Using a single spatial wavefront to control polychromatic light does introduce a loss of control, on top of the incomplete channel control~\cite{2013_Goetschy_PRL} present in previous WFS experiments with narrowband light.
But long-range spectral correlations significantly reduce this loss, making the coherent control of short pulses and multi-frequency laser beams potentially feasible despite the rather narrow spectral correlation width, $\delta \omega$, of the open channels and CEA channels.
The formalism for uncorrelated matrices (Eq.~\eqref{eq:gA}) can also treat spatially incoherent light with multiple uncorrelated transverse modes or unpolarized light with two independent polarizations.

We acknowledge helpful discussions with Seng Fatt Liew,  Alexey Yamilov, Raktim Sarma, and Steven G Johnson.
This work is supported by the National Science Foundation under grant No.~DMR-1307632, DMR-1205307, and  ECCS-1068642, and by the US Office of Naval Research under grant No.~N00014-13-1-0649. 

\bibliography{mybib}

\begin{thebibliography}{39}
\bibitem[{not({\natexlab{b}})}]{note_CEA}
\bibinfo{note}{One can interpret the $N^2 l/l_a \gg 1$ condition heuristically:
  strong absorption requires penetration depth of $L_a$, and in a low-loss
  medium, existence of high-transmission channels into depth $L_a$ requires $g
  = N l/L_a \gg 1$.}
\end{thebibliography}

\end{document}


\title{Supplementary Material}
\maketitle

\renewcommand{\theequation}{S.\arabic{equation}}
\setcounter{equation}{0}

\renewcommand{\thefigure}{S\arabic{figure}}
\setcounter{figure}{0}

\subsection{Eigenvalue Distribution for Sums of Uncorrelated Matrices}

\noindent In this section, we provide details for obtaining the eigenvalue distribution for a sum of uncorrelated matrices, namely Eq.~(2) in the main text and its extensions. For a Hermitian matrix $X$ with eigenvalue distribution $p_X$, its Stieltjes transform (also called resolvent) $g_X$ is~\cite{Voiculescu_book, Tulino_book}
\begin{equation}
\label{eq:resolvent}
g_X(z) = \int_{-\infty}^{\infty}  \frac{p_X(z')}{z - z'} dz'
\end{equation}
for complex-valued $z$.
For example, for a transmission matrix $t_m$ in the diffusive regime ($\lambda \ll l \ll L$), the eigenvalue distribution of $t_m^\dagger t_m$ follows~\cite{1994_Nazarov_PRL}
\begin{align}
\label{eq:p0_T}
p_{t_m^\dagger t_m}(T) = \left\{ 
\begin{aligned}
& \frac{\bar{T}_m}{2T\sqrt{1-T}} ,& {\rm sech}^2(\frac{1}{\bar{T}_m}) \le T < 1 \\
& 0 ,& {\rm otherwise}
\end{aligned}
\right.
\end{align}
where $\bar{T}_m$ is the average transmission, so its Stieltjes transform is
\begin{equation}
\label{eq:g0_T}
g_{t_m^\dagger t_m}(z) = \frac{1}{z} - \frac{\bar{T}_m}{z\sqrt{1-z}}{\rm Arctanh}\left[\frac{{\rm Tanh}(1/\bar{T}_m)}{\sqrt{1-z}} \right].
\end{equation}
For the reflection matrix $r_m$ in the diffusive regime with weak absorption that we consider in CEA ($\lambda \ll l \ll \sqrt{l l_a} < L$),  the eigenvalue distribution of $r_m^\dagger r_m$ follows~\cite{1996_Bruce_JPA}
\begin{align}
\label{eq:p0_R}
p_{r_m^\dagger r_m}(R) = \left\{ 
\begin{aligned}
& \frac{2a_m}{\pi(1-R)^2} \sqrt{\frac{1-R}{a_mR} - 1},& 0 < R \le  \frac{1}{1+a_m} \\
& 0 ,& {\rm otherwise}
\end{aligned}
\right.
\end{align}
where $a_m = l/l_a = (1-\bar{R}_m)^2/4\bar{R}_m$,  so its Stieltjes transform is
\begin{equation}
\label{eq:g0_R}
g_{r_m^\dagger r_m}(z) = \frac{z - 1 + 2a_m  - 2a_m \sqrt{1 + \frac{1}{a_m} - \frac{1}{a_mz}}}{(1-z)^2}.
\end{equation}
Given the Stieltjes transform, we can apply Cauchy's integral formula to get back the eigenvalue distribution
\begin{equation}
\label{eq:g2p}
p_X(z) = - \frac{1}{\pi} \lim_{\epsilon \to 0^+} {\rm Im} g_X(z+i\epsilon).
\end{equation}

In general, knowing the eigenvalue distribution of two individual matrices is not enough to infer the eigenvalue distribution of their sum, as the two matrices do not share the same eigenbasis.
However, free probability theory identifies a sufficient condition, called asymptotic freeness, under which the eigenvalue distribution of the summed matrix cab be computed.
When $X_1$ and $X_2$ are asymptotically free, their $R$-transforms are additive~\cite{Voiculescu_book, Tulino_book, 1986_Voiculescu_addition}
\begin{equation}
\label{eq:addition_rule}
R_{X_1 + X_2}(y) = R_{X_1}(y) + R_{X_2}(y),
\end{equation}
where the $R$-transform is defined as
\begin{equation}
R_X(y) = g_X^{-1}(y) - \frac{1}{y},
\end{equation}
with $g_X^{-1}$ being the functional inverse of $g_X$.

For a discrete spectrum $I(\omega) = \sum_m W_m \delta(\omega-\omega_m)$, Eq.~(1) of the main text becomes
\begin{equation}
\label{eq:A_discrete}
A = \sum_{m} W_m t_m^\dagger t_m,
\end{equation}
where $t_m = t(\omega_m)$.
In general, two Hermitian random matrices of size $N$ are asymptotically free in the limit $N \to \infty$ if the two matrices are uncorrelated to each other. This is the case for the set of matrices $ \{W_m t_m^\dagger t_m\}$ in Eq.~\eqref{eq:A_discrete} if the frequencies $\{\omega_m\}$ are much further than $\delta \omega$ apart.
Applying Eq.~\eqref{eq:addition_rule} to Eq.~\eqref{eq:A_discrete} yields
\begin{equation}
\label{eq:gA_SI}
z - \frac{1}{g_A(z)} = \sum_{m} \left[ W_m g_{t_m^\dagger t_m}^{-1}(W_m g_A(z)) -  \frac{1}{g_A(z)}  \right],
\end{equation}
which is Eq.~(2) in the main text.
Given the weights $\{W_m\}$ and the average transmissions $\{\bar{T}_m\}$ (or $\{\bar{R}_m\}$ for the reflection case), we can numerically solve Eq.~\eqref{eq:gA_SI} for $g_A(z)$ using the known $g_{t_m^\dagger t_m}$ from Eq.~\eqref{eq:g0_T} (or $g_{r_m^\dagger r_m}$ from Eq.~\eqref{eq:g0_R}), and obtain the eigenvalue distribution $p_A$ through Eq.~\eqref{eq:g2p}.

The minimal and the maximal eigenvalues also follow from Eq.~\eqref{eq:gA_SI}.
At the lower or upper edge $z_0$ of the distribution, the probability density becomes zero, and its slope diverges. 
That means $g_A(z_0)$ is purely real, and $|g'_A(z_0)| = \infty$.
Taking the $z$ derivative on both sides of Eq.~\eqref{eq:gA_SI}, we see that $|g'_A(z_0)| = \infty$ when
\begin{equation}
\label{eq:edge}
\frac{1}{g_A^2(z_0)} = \sum_{m} \left[ \frac{W_m^2}{ g_{t_m^\dagger t_m}'\left(g_{t_m^\dagger t_m}^{-1} \left(W_m g_A \left(z_0\right)\right)\right) } + \frac{1}{g_A^2(z_0)} \right] .
\end{equation}
We can solve Eq.~\eqref{eq:edge} for the real-valued $g_A(z_0)$, and plug $g_A(z_0)$ back into Eq.~\eqref{eq:gA_SI} to obtain $z_0$.
In general, there will be two solutions of $g_A(z_0)$, which gives the minimal and the maximal eigenvalues.

The unknown in Eq.~\eqref{eq:edge} is real-valued and can be obtained with basic bracketing methods.
When one edge $z_0$ is found, we can start from $z=z_0$ with the known $g_A(z_0)$ being the initial guess, and use Newton's method to solve Eq.~\eqref{eq:gA_SI} for the complex-valued $g_A(z)$, with $z$ changing incrementally based on the magnitude of the Jacobian.

\begin{figure}[t]
   \includegraphics{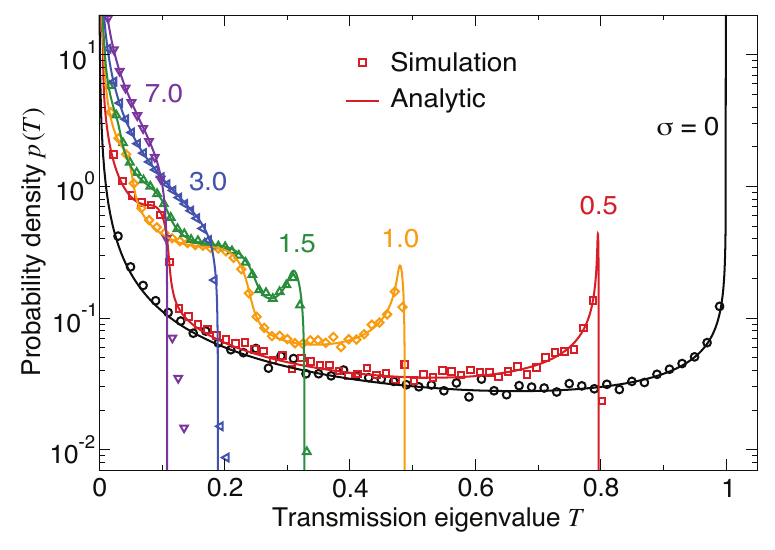} 
   \caption{Distribution of transmission eigenvalues for a set of discrete and uncorrelated frequencies with weights following a discrete Gaussian distribution with standard deviation $\sigma$. Symbols: simulation. Lines: solution of Eq.~\eqref{eq:gA_SI}.
}
   \label{fig:pT_gaussian}
\end{figure}

In Fig.~\ref{fig:pT_gaussian}, we plot the solution of Eq.~\eqref{eq:gA_SI} with the weights $\{W_m\}$ given by a discrete Gaussian distribution with standard deviation $\sigma$.
To validate the analytical prediction, we carry out wave simulations of diffusive media (see main text for system parameters; here $\omega = 410 c/L$, $\bar{T} = 0.021$), with the uncorrelated matrices given by the transmission matrices of different disorder realizations. The numerically calculated eigenvalue densities are shown as symbols in Fig.~\ref{fig:pT_gaussian} and agree perfectly with the analytic prediction with no fitting parameters.

The effects of incomplete channel control, common in experiments, can be included in our formalism by replacing $t_m$ and $r_m$ with the filtered matrices $\tilde{t}_m$ and $\tilde{r}_m$ introduced in Ref.~\cite{2013_Goetschy_PRL}.

\subsection{Spectral Correlation Width for Transmission}

\begin{figure}[t]
   \includegraphics{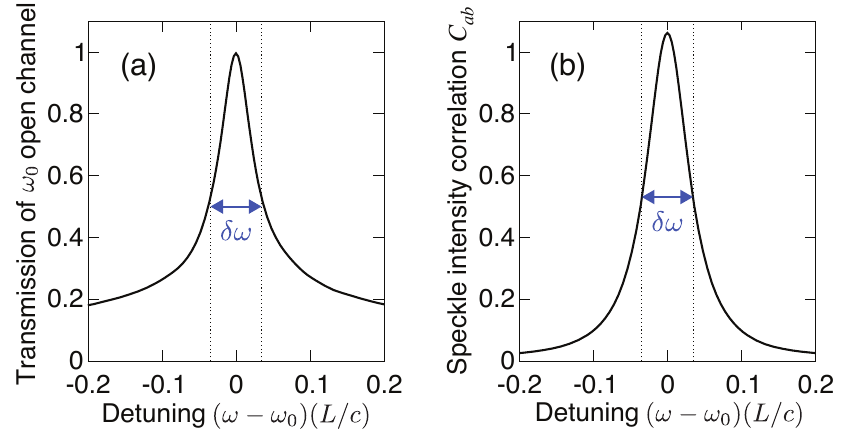} 
   \caption{Spectral correlation of transmission through disordered media, for the lossless system considered in Figs.~1-3 of the main text.
   (a) Transmission spectrum for the monochromatic transmission eigenchannel with the highest transmission at $\omega_0$.
   (b) Spectral correlation function of the speckle intensity.
}
   \label{fig:dw_T}
\end{figure}

\noindent In Fig.~\ref{fig:dw_T}(a), we show the transmission spectrum of the monochromatic open channel, for the lossless system considered in the main text; its full width at half maximum (FWHM) is what we define as the spectral correlation width $\delta \omega$. As shown in Fig.~\ref{fig:dw_T}(b), this $\delta \omega$ coincides with the FWHM of the spectral correlation function of the speckle intensity, $C_{ab}(\omega_1, \omega_2) \equiv {\langle T_{ab}(\omega_1) T_{ab}(\omega_2)\rangle}/{\langle T_{ab}(\omega_1)\rangle \langle T_{ab}(\omega_2)\rangle} - 1$, where $T_{ab} = |t_{ba}|^2$.

\subsection{Relating $M_{\rm eff}$ to Spectral Correlation}

\noindent Here we derive and validate Eq.~(5) in the main text and its more general form, which shows that the effective number $M_{\rm eff}$ of uncorrelated frequencies is simply the inverse of the frequency-averaged spectral correlation of the total transmission or total reflection.
To compute $M_{\rm eff} = {\rm Var}({\tau})/{\rm Var}(\tilde{\tau})$ as defined in Eq.~(4) of the main text, we need to evaluate
\begin{align}
\label{eq:var_tau}
\begin{split}
&{\rm Var}(\tilde{\tau})
= \frac{1}{N} \langle {\rm Tr} (A A) \rangle - \left(  \frac{1}{N} \langle {\rm Tr} (A ) \rangle \right)^2 \\
& \qquad= \iint d\omega_1 d\omega_2 I(\omega_1) I(\omega_2)
\bar{T}_1 \bar{T}_2 F^{(T)}(\omega_1, \omega_2)
\end{split}
\end{align}
where
\begin{equation}
\label{eq:F_def}
F^{(T)}(\omega_1, \omega_2) \equiv  \frac{ \langle {\rm Tr} (t_1^\dagger t_1 t_2^\dagger t_2) \rangle}{N \bar{T}_1 \bar{T}_2 } - 1,
\end{equation}
$t_i = t(\omega_i)$, $\bar{T}_i = \langle {\rm Tr} (t_i^\dagger t_i) \rangle/N$, and the brackets denote averaging over different disordered samples.

\begin{figure*}[t]
   \includegraphics{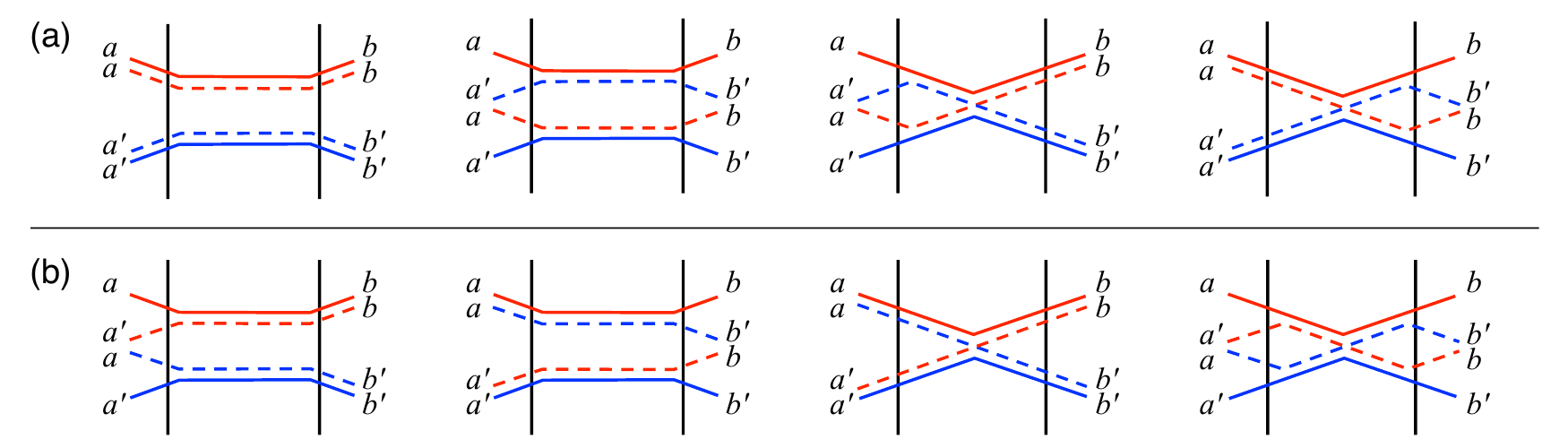} 
   \caption{
   Diagrams showing the leading-order contributions for
   (a) the term $\langle t_{ba}(\omega_1)t_{ba}^*(\omega_1)t_{b'a'}(\omega_2)t_{b'a'}^*(\omega_2) \rangle$ in the speckle intensity correlation, and
   (b) the term $\langle t_{ba}(\omega_1)t_{ba'}^*(\omega_1)t_{b'a'}(\omega_2)t_{b'a}^*(\omega_2) \rangle$ in the trace in the definition of $F^{(T)}$.
Red and blue colors indicate propagation at $\omega_1$ and $\omega_2$ respectively.
Solid and dashed lines correspond to retarded Green's function (unconjugated transmission coefficient) and advanced Green's function (conjugated transmission coefficient), respectively.
Vertical black lines indicate the boundary of the disordered medium.
}
   \label{fig:trace_schematic}
\end{figure*}

The quantity $F^{(T)}$ contains
$\langle {\rm Tr} (t_1^\dagger t_1 t_2^\dagger t_2) \rangle = \sum_{a,b,a',b'} \langle t_{ba}(\omega_1)t_{ba'}^*(\omega_1)t_{b'a'}(\omega_2)t_{b'a}^*(\omega_2) \rangle$.
Meanwhile, the spectral-and-channel correlation of the speckle intensity involves a closely related quantity
$\langle T_{ab}(\omega_1) T_{a'b'}(\omega_2)\rangle =  \langle t_{ba}(\omega_1)t_{ba}^*(\omega_1)t_{b'a'}(\omega_2)t_{b'a'}^*(\omega_2) \rangle$,
which in the diffusive regime 
separates into three contributions, $C_1$, $C_2$, and $C_3$~\cite{1988_Feng_PRL, 1988_Mello_PRL} (also see reviews in Refs.~\cite{1994_Berkovits_PR, 2001_Pnini_chapter, Akkermans_book, Sheng_book}).
In waveguide geometries where the incident and outgoing channels are quantized into waveguide modes, the channel dependences appear as Kronecker delta symbols~\cite{1988_Mello_PRL, 1991_Mello_Stone_PRB},
and the $C_1$ and $C_2$ contributions of $\langle T_{ab}(\omega_1) T_{a'b'}(\omega_2)\rangle$, shown diagrammatically in Fig.~\ref{fig:trace_schematic}(a), give
\begin{align}
\label{eq:intensity_correlation}
\begin{split}
& \langle t_{ba}(\omega_1)t_{ba}^*(\omega_1)t_{b'a'}(\omega_2)t_{b'a'}^*(\omega_2) \rangle
= \langle T_{ab}\rangle \langle T_{a'b'} \rangle \times \\
& \qquad \left[ 1 + \delta_{aa'}\delta_{bb'}C_1(\omega_1, \omega_2) + (\delta_{aa'} + \delta_{bb'}) C_2(\omega_1, \omega_2) \right].
\end{split}
\end{align}
The $C_3$ contribution is negligible for our purpose here.
Also, we drop the weak frequency dependence of $\langle T_{ab}\rangle$ and $\langle T_{a'b'} \rangle$. 
The four-amplitude average in the trace of Eq.~\eqref{eq:F_def} differs from Eq.~\eqref{eq:intensity_correlation} only by switching the indices $a$ and $a'$ on the conjugated amplitudes; diagrammatically this corresponds to bending the legs of the dashed lines on the incident side, as shown in  Fig.~\ref{fig:trace_schematic}(b),
and the effect is to swap $\delta_{aa'}$ and 1 in each term, giving
\begin{align}
\label{eq:trace_correlation}
\begin{split}
& \langle t_{ba}(\omega_1)t_{ba'}^*(\omega_1)t_{b'a'}(\omega_2)t_{b'a}^*(\omega_2) \rangle
= \langle T_{ab}\rangle \langle T_{a'b'} \rangle \times \\
& \qquad \left[ \delta_{aa'}  +  \delta_{bb'} C_1(\omega_1, \omega_2) + (1+ \delta_{aa'}\delta_{bb'}) C_2(\omega_1, \omega_2) \right].
\end{split}
\end{align}
Summing over $a, a', b, b'$, the fourth term becomes negligible, and we get
$F^{(T)}(\omega_1, \omega_2) =  C_1(\omega_1, \omega_2) + N C_2(\omega_1, \omega_2)$.
Instead of $C_2$, 
it is easier to use the mean-normalized correlation $C^{(T)}(\omega_1,\omega_2) \equiv \langle T_a(\omega_1) T_a(\omega_2) \rangle / \langle T_a(\omega_1)\rangle \langle T_a(\omega_2) \rangle - 1$ of the total transmission  $T_a = \sum_b |t_{ba}|^2$.
Setting $a=a'$ and summing over $b$ and $b'$ in Eq.~\eqref{eq:intensity_correlation}, we see that 
 $C^{(T)}(\omega_1,\omega_2) = (1/N) C_1(\omega_1, \omega_2) + C_2(\omega_1, \omega_2)$, so we get
\begin{equation}
\label{eq:F_and_C2T}
F^{(T)}(\omega_1, \omega_2) = N C^{(T)}(\omega_1,\omega_2).
\end{equation}

\begin{figure}[t]
   \includegraphics{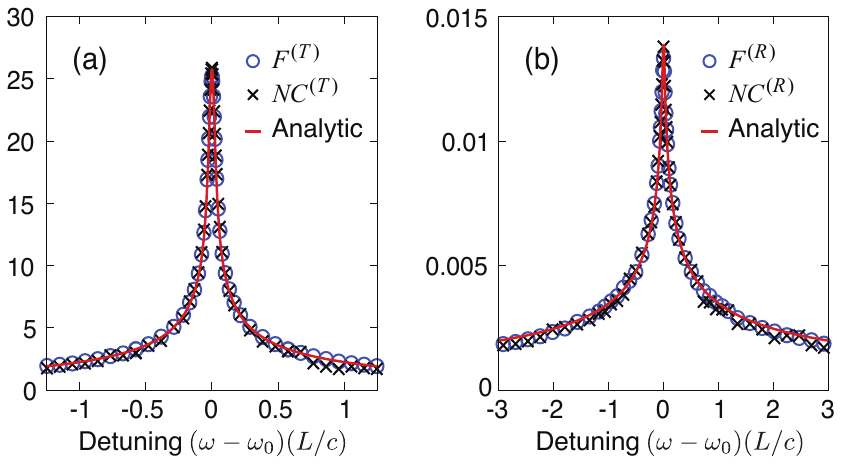} 
   \caption{Numerical validation of $F^{(T,R)} = N C^{(T,R)}$ and of the analytical expressions for $C^{(T,R)}$,
   for transmission in the lossless system considered in Figs.~1-3 of the main text (a)
   and reflection in the weakly absorbing system considered in Fig.~4 of the main text (b).
   Blue circles are numerically calculated $F^{(T)}$ and $F^{(R)}$ as defined in Eq.~\eqref{eq:F_def}.
   Black crosses are numerically calculated spectral correlation of the total transmission and total reflection, multiplied by the number of modes $N$.
   Red solid lines are the analytical expressions in Eq.~\eqref{eq:C_T_SI} and Eq.~\eqref{eq:C_R_SI}.
}
   \label{fig:F_and_C}
\end{figure}

To test the validity of Eq.~\eqref{eq:F_and_C2T}, in Fig.~\ref{fig:F_and_C}(a) we plot $F^{(T)}$ (blue circles) and $N C^{(T)}$ (black crosses) calculated from numerically obtained transmission matrices of the lossless system considered in the main text.
The agreement remains excellent even when absorption is included (data not shown).

Inserting Eq.~\eqref{eq:F_and_C2T} into Eq.~\eqref{eq:var_tau} and dropping the weak frequency dependence of $\bar{T}$, we get
\begin{equation}
\label{eq:var_and_C}
\frac{{\rm Var}(\tilde{\tau}) }{{\rm Var}({\tau})} = \iint d\omega_1 d\omega_2 I(\omega_1) I(\omega_2) \frac{C^{(T)}(\omega_1,\omega_2)}{C^{(T)}(\omega_0, \omega_0)}.
\end{equation}
Eq.~\eqref{eq:var_and_C} holds for arbitrary spectrum $I(\omega)$ as long as $\bar{T}$ is roughly constant within the bandwidth and that the system is in the diffusive regime.
However, the relation $M_{\rm eff} = {\rm Var}({\tau})/{\rm Var}(\tilde{\tau})$ is only meaningful when the eigenvalue density of the broadband matrix $A$ follows the equal-weight uncorrelated model (Eq.~(3) in the main text), which is not necessarily the case when the spectrum $I(\omega)$ exhibits complex structures.
When $I(\omega)$ is a smooth function without large variation, the equal-weight uncorrelated model works well, and we can use Eq.~\eqref{eq:var_and_C} to obtain $M_{\rm eff}$.
Eq.~(5) in the main text is the special case when $I(\omega)$ is a uniform distribution across bandwidth $\Delta \omega$.

Similarly, for the reflection matrices, we have
\begin{equation}
\label{eq:F_and_C2R}
F^{(R)}(\omega_1, \omega_2) = N C^{(R)}(\omega_1,\omega_2),
\end{equation}
where 
$F^{(R)}(\omega_1, \omega_2) \equiv { \langle {\rm Tr} (r_1^\dagger r_1 r_2^\dagger r_2) \rangle}/({N \bar{R}_1 \bar{R}_2 }) - 1$, and
$C^{(R)}(\omega_1,\omega_2) \equiv \langle R_a(\omega_1) R_a(\omega_2) \rangle / \langle R_a(\omega_1)\rangle \langle R_a(\omega_2) \rangle - 1$ with the total reflection being $R_a = \sum_b |r_{ba}|^2$.
Although the reflection correlation is more complicated involving the effect of coherent back scattering~\cite{1989_Wang_PRB, 1990_Berkovits_PRB_2} and of short optical paths~\cite{1995_Rogozkin_PRB}, these effects are present in both $F^{(R)}$ and $C^{(R)}$, so the final expression Eq.~\eqref{eq:F_and_C2R} appears to be unchanged.
Indeed, in Fig.~\ref{fig:F_and_C}(b) we numerically validate Eq.~\eqref{eq:F_and_C2R} for the weakly absorbing system considered in the main text, and perfect agreement is observed.
Eq.~\eqref{eq:F_and_C2R} also holds in the absence of absorption (data not shown).
Therefore, Eq.~(5) in the main text is also valid for reflection, just with $C^{(T)}$ replaced by $C^{(R)}$.

In the absence of absorption, we can derive Eqs.~\eqref{eq:F_and_C2R} from \eqref{eq:F_and_C2T} or the other way around based only on energy conservation.
Energy conservation requires that $r^\dagger r + t^\dagger t = \mathbb{I}$ at both frequencies, so
$\langle {\rm Tr} (t_1^\dagger t_1 t_2^\dagger t_2) \rangle/N -  \bar{T}_1 \bar{T}_2 
= \langle {\rm Tr} (r_1^\dagger r_1 r_2^\dagger r_2) \rangle/N -  \bar{R}_1 \bar{R}_2$.
Meanwhile, $R_a(\omega)+T_a(\omega)=1$ for every incident channel $a$, so
$ \bar{T}_1 \bar{T}_2 C^{(T)}(\omega_1, \omega_2) = \bar{R}_1 \bar{R}_2 C^{(R)}(\omega_1, \omega_2)$.
These two requirements demand Eq.~\eqref{eq:F_and_C2R} to follow from Eq.~\eqref{eq:F_and_C2T}.

When $\omega_1 = \omega_2$, the trace in $F^{(T,R)}$ is simply the second moment of the eigenvalues, which we can evaluate directly from Eq.~\eqref{eq:p0_T} and Eq.~\eqref{eq:p0_R}.
In such case, Eqs.~\eqref{eq:F_and_C2T}-\eqref{eq:F_and_C2R} yields
\begin{align}
\label{eq:CTR_0}
\begin{split}
C^{(T)}(\omega_0, \omega_0) &= \frac{1}{N\bar{T}} \left( \frac{2}{3} - \bar{T} \right),  \\
C^{(R)}(\omega_0, \omega_0) &= \frac{1-\bar{R}}{N(1+\bar{R})},
\end{split}
\end{align}
for the systems of interest here. We will use these equalities for the analytical expressions of $C^{(T)}$ and $C^{(R)}$ in the next two sections.

We also note that when $\omega_1 = \omega_2$, Eqs.~\eqref{eq:F_and_C2T}
reduces to a simple statement ${\rm Var}(\tau) = N {\rm Var}(T_a)$ that is a direct consequence of the randomness of the matrices and the small fluctuations of the eigenvalues with respect to disorder realizations.
The transmission matrix $t$ has singular value decomposition $t = U \tau^{1/2} U^\dagger$, where $\tau$ is a diagonal matrix of the transmission eigenvalues, and $U$, $V$ are unitary matrices.
Then, $\langle T_a \rangle = \sum_n \langle  |V_{an}|^2 \tau_n \rangle$ and 
$\langle T_a^2 \rangle = \sum_{n,m} \langle  |V_{an}|^2 |V_{am}|^2 \tau_n \tau_m \rangle$.
The eigenvalues  $\{\tau_n\}$ are not correlated with the matrix $V$, and the sum of eigenvalues ($N\bar{T}$) has a fluctuation with respect to disorder that is of order unity~\cite{1985_Lee_Stone_PRL} and therefore negligible when $N\bar{T}$ is large, so it suffices to take the ensemble average over the unitary matrix $V$.
We further invoke the isotropic approximation that all incident channels $a$ are equivalent, so the ensemble average becomes the average over the invariant measure of the unitary group, for which
 $\langle  |V_{an}|^2 \rangle = 1/N$ and $\langle  |V_{an}|^2 |V_{am}|^2 \rangle = (1+\delta_{nm})/(N(N+1))$ ({\it e.g.}, see Refs.~\cite{1990_Mello_JPA, Tulino_book}).
This gives ${\rm Var}(T_a) = \langle T_a^2 \rangle  - \langle T_a \rangle^2 = (\langle \tau^2 \rangle - \bar{T}^2)/(N+1) = {\rm Var}(\tau)/(N+1)$.
Note that here ${\rm Var}(T_a)$ is the variation with respect to disorder realizations (while $a$ is fixed), whereas ${\rm Var}(\tau)$ is the variation with respect to the eigenvalue distribution. 
This derivation also applies to the reflection matrices.

\subsection{Analytical Expression for Transmission $C^{(T)}$}

\noindent Refs.~\cite{1987_Stephen_PRL, 1989_Pnini_PRB} derived the spectral correlation of the total transmission, as
\begin{equation}
\label{eq:CT_nocorrection}
C^{(T)}(\omega_1, \omega_2) =  \frac{1}{N\bar{T}}\frac{2}{x} \frac{\sinh(x) - \sin(x)}{\cosh(x) - \cos(x)},
\end{equation}
where $x = \sqrt{2 {|\omega_1 - \omega_2|}L^2/{D}}$, $L$ is the sample thickness, and $D$ is the diffusion coefficient.
For large frequency difference, $C^{(T)}$ has a long-range $|\omega_1-\omega_2|^{-1/2}$ dependence.
Various corrections has been derived to account for the effects of absorption~\cite{1991_Pnini_PLA, 1992_Kogan_PRB}, finite illumination area~\cite{1992_deBoer_PRB}, and interfacial layer~\cite{1993_Lisyansky_PRB, 1993_vanRossum_PLA}.
However, Eq.~\eqref{eq:CT_nocorrection} predicts that $C^{(T)}(\omega_0, \omega_0) = (2/3)/(N\bar{T})$, which differs from Eq.~\eqref{eq:CTR_0} by a shift that vanishes with $\bar{T}$.
Indeed, from the numerically calculated transmission matrices, we find that Eq.~\eqref{eq:CT_nocorrection} differs from the numerical results by a constant shift that vanishes when $\bar{T} \to 0$.
Such a constant shift was observed experimentally in Fig.~5 of Ref.~\cite{1992_deBoer_PRB}, although it was treated as an artifact of the averaging procedure.
Empirically, we find that the following expression accounts for the constant shift and yields the correct value of $C^{(T)}(\omega_0, \omega_0)$,
\begin{equation}
\label{eq:C_T_SI}
C^{(T)}(\omega_1, \omega_2) = \frac{1}{N\bar{T}} \left[ \frac{2}{x} \frac{\sinh(x) - \sin(x)}{\cosh(x) - \cos(x)} - \bar{T} \right].
\end{equation}
This is Eq.~(6) in the main text.
This expression is plotted as the red solid line in Fig.~\ref{fig:F_and_C}(a). 
The only unknown parameter in Eq.~\eqref{eq:C_T_SI} is the diffusion coefficient, which we set to $D = 3.3\times 10^{-3} c L$.
Once $D$ is known from the the transmission correlation, we can use Eq.~(5) in the main text to obtain $M_{\rm eff}$ analytically.

\subsection{Analytical Expression for Reflection $C^{(R)}$}

\noindent Rogozkin and Cherkasov~\cite{1995_Rogozkin_PRB} derived the spectral correlation of the total reflection for a scattering medium with weak absorption ($a = \sqrt{l/l_a} \to 0$).
Taking the $L \gg \sqrt{l l_a}$ limit of Eqs.~(B5a) and (B6) in Ref.~\cite{1995_Rogozkin_PRB}, we obtain
\begin{equation}
\label{eq:C2_Rogozkin}
C^{(R)}(\omega_1, \omega_2) = \frac{1-\bar{R}}{N} \frac{1}{1+y} ,
\end{equation}
where $y={\rm Re}\sqrt{1+ i |\omega_1-\omega_2| l_a /c}$.
As in the transmission case, Eq.~\eqref{eq:C2_Rogozkin} exhibits long-range $|\omega_1-\omega_2|^{-1/2}$ dependence for large frequency difference.
However, Eq.~\eqref{eq:C2_Rogozkin} was derived for very weak absorption ($a \to 0$, and $\bar{R} \to 1$), and it only agrees with Eq.~\eqref{eq:CTR_0} in this limit.
Indeed, from the numerically calculated reflection matrices, we find Eq.~\eqref{eq:C2_Rogozkin} to differ from the numerical results by a constant shift that vanishes when $\bar{R} \to 1$, similar to the transmission case.
Empirically, we find that the following expression accounts for the constant shift and yields the correct value of $C^{(R)}(\omega_0, \omega_0)$,
\begin{equation}
\label{eq:C_R_SI}
C_2^{\rm (R)}(\omega_1, \omega_2) = \frac{1-\bar{R}}{N(1+\bar{R})\bar{R}} \left[ \frac{2}{1+y} - (1-\bar{R}) \right].
\end{equation}
This is Eq.~(7) in the main text.
This expression is plotted as the red solid line in Fig.~\ref{fig:F_and_C}(b).
The only unknown parameter in Eq.~\eqref{eq:C_R_SI} is the ballistic absorption length, which we set to $l_a = 83 L$.
Once $l_a$ is known from the the reflection correlation, we can use Eq.~(5) in the main text to obtain $M_{\rm eff}$ analytically.

\begin{figure}[t]
   \includegraphics{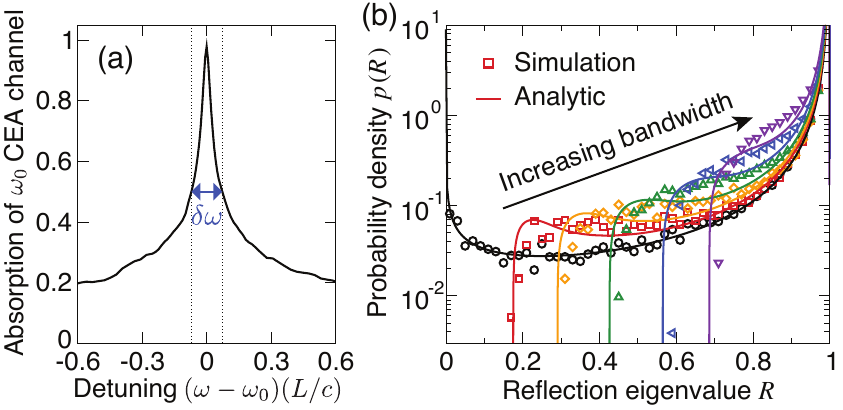} 
   \caption{
   (a) Absorption spectrum for the monochromatic reflection eigenchannel with the highest absorption at $\omega_0$.
   (b) Density of polychromatic reflection eigenvalues for various bandwidths, calculated numerically from simulations (symbols; bandwidths $\Delta \omega / \delta \omega =$ 0, 1.2, 2.7, 6.6, 17, 45) and analytically from Eq.~(3) of the main text with an effective number $M_{\rm eff}$ of uncorrelated frequencies (lines; $M_{\rm eff} = 1.0, 1.3, 1.6, 2.0, 2.9, 4.4$).
}
   \label{fig:dw_pR_CEA}
\end{figure}

\subsection{Broadband Coherently Enhanced Absorption}

\noindent Here we provide additional data for the weakly absorbing system considered in Fig.~4 of the main text for broadband CEA.
Fig.~\ref{fig:dw_pR_CEA}(a) shows the absorption spectrum for the monochromatic reflection eigenchannel with the highest absorption (lowest reflection); its FWHM is what we use to define $\delta \omega$.
In Fig.~\ref{fig:dw_pR_CEA}(b), we plot the distribution of broadband reflection eigenvalues for various bandwidths $\Delta \omega$, comparing the simulation results (symbols) and the prediction from the uncorrelated model (lines) with an effective number $M_{\rm eff}$ of uncorrelated frequencies.

\bibliography{mybib}